\documentclass[conference]{IEEEtran}
\IEEEoverridecommandlockouts

\usepackage[T1]{fontenc}
\usepackage{amsmath,amssymb,amsfonts}
\usepackage{graphicx}
\usepackage{cite}

\usepackage{array}
\usepackage{caption}
\usepackage{enumitem}
\usepackage{float}
\usepackage{algorithmic}
\usepackage{textcomp}

\usepackage{multirow}
\usepackage{longtable}
\usepackage{colortbl}

\usepackage{xcolor}
\usepackage{tikz}
\usepackage{pgfplots}
\usepackage{svg}

\usetikzlibrary{shapes.geometric, arrows, positioning, calc}

\pgfplotsset{compat=1.18}

\definecolor{coldblue}{RGB}{135,206,235}
\definecolor{freezingblue}{RGB}{0,127,255}
\definecolor{hotred}{RGB}{255,160,122}
\definecolor{scorchingred}{RGB}{255,69,0}
\definecolor{warmyellow}{RGB}{255,215,0}
\definecolor{negligiblegray}{RGB}{220,220,220}

\usepackage{hyperref}
\hypersetup{
    colorlinks   = true,
    urlcolor     = blue,
    linkcolor    = blue,
    citecolor    = blue
}

\def\BibTeX{{\rm B\kern-.05em{\sc i\kern-.025em b}\kern-.08em
    T\kern-.1667em\lower.7ex\hbox{E}\kern-.125emX}}

\begin{document}
\title{BioCrypt : Gradient-based facial encoding for key generation to encrypt and decrypt multimedia data\\
\vspace{0.5cm}
}
\author{%
    \IEEEauthorblockN{1\textsuperscript{st} Ankit Kumar Patel}
    \IEEEauthorblockA{\textit{B.Tech Electronics \& Computer Engineering} \\
        \textit{Graduated from VIT University}\\
        Chennai, India \\
        rklankit12@gmail.com}
    \and
    \IEEEauthorblockN{2\textsuperscript{nd} Dewanshi Paul}
    \IEEEauthorblockA{\textit{B.Tech Electronics \& Computer Engineering} \\
        \textit{Graduated from VIT University}\\
        Chennai, India \\
        dewanshipaul84@gmail.com}
    \and
    \IEEEauthorblockN{3\textsuperscript{rd} Sarthak Giri}
    \IEEEauthorblockA{\textit{B.Tech Computer Science \& Engineering} \\
        \textit{Graduated from VIT University}\\
        Vellore, India \\
        girisarthak84@gmail.com}
    \and
    \IEEEauthorblockN{4\textsuperscript{th} Sneha Chaudhary}
    \IEEEauthorblockA{\textit{B.Tech Computer Science \& Engineering} \\
        \textit{Graduated from VIT University}\\
        Vellore, India \\
        chaudhary.sneha544@gmail.com}
    \and
    \IEEEauthorblockN{5\textsuperscript{th} Bikalpa Gautam}
    \IEEEauthorblockA{\textit{B.Tech Civil Engineering} \\
        \textit{Graduated from VIT University}\\
        Vellore, India \\
        bikalpa.gautam111@gmail.com}
}
\maketitle
\vspace{-0.5cm}
\renewcommand{\thefootnote}{\textit{}} 

\footnotetext{  
\hspace{-0.365cm} \rule{\textwidth}{0.4pt}  
The following abbreviations and symbols are used in this manuscript:\vspace{0.12cm}\\
 \begin{tabular*}{\textwidth}{@{\extracolsep{\fill}}ll@{}ll@{}ll@{}}
 \textbf{Abbreviations} & & \textbf{Symbols} & & \textbf{Symbols} & \\
 AES & Advanced Encryption Standard & $C$ & Ciphertext & $r_{xy}$ & Correlation coefficient \\
 HOG & Histogram of Oriented Gradients & $P$ & Plaintext & $H(X)$ & Shannon entropy \\
 SVM & Support Vector Machines & $IV$ & Initialization Vector & $dnorm$ & Normalized Hamming Distance \\
 CBC & Cipher Block Chaining & $n$ & Total number of bits & $d$ & Number of differentiating bits \\
 ROC & Receiver Operating Characteristics & $R$ & Sample rate (Hz) & $\oplus$ & XOR operation \\
 MP3 & MPEG Audio Layer-3 & $F$ & Feature dimension & $H_{DW}(\tau)$ & MCCC weighting function \\
 MP4 & MPEG-4 Part 14 & $Acc$ & Accuracy & $\log_b(p(x_i))$ & Logarithmic probability\\
 TIFF & Tagged Image File Format & $L_P$ & Length of plaintext & $C_i$ & Encrypted data block\\
 PDF & Portable Document Format & $F_1$ & F1-Score & $P_i$ & Decrypted plaintext block\\
 PPTX & PowerPoint Open XML Presentation & $tr(R)$ & Trace of matrix $R$ & $\Psi(f)$ & PHAT function\\
 FLV & Flash Video & $B$ & Batch size & $\Delta P$ & Change in plaintext\\
 IPYNB & Interactive Python Notebook & $EER$ & Equal Error Rate & $\eta$ & Random initialization entropy\\
 \end{tabular*}
}

\begin{abstract}
Security systems relying on passwords are vulnerable to being forgotten, guessed, or breached. Likewise, biometric systems that operate independently are at risk of template spoofing and replay incidents. This paper introduces a biocryptosystem utilizing face recognition techniques to address these issues, allowing for the encryption and decryption of various file types through the Advanced Encryption Standard (AES). The proposed system creates a distinct 32-bit encryption key derived from facial features identified by Histogram of Oriented Gradients (HOG) and categorized using Support Vector Machines (SVM). HOG efficiently identifies edge-aligned facial features, even in dim lighting, ensuring that reliable biometric keys can be generated. This key is then used with AES to encrypt and decrypt a variety of data formats, such as text, audio, and video files. This encryption key, derived from an individual's distinctive facial traits, is exceedingly challenging for adversaries to reproduce or guess. The security and performance of the system have been validated through experiments using several metrics, including correlation analysis, Shannon entropy, normalized Hamming distance, and the avalanche effect on 25 different file types. Potential uses for the proposed system include secure file sharing, online transactions, and data archiving, making it a strong and trustworthy approach to safeguarding sensitive information by integrating the uniqueness of facial biometrics with the established security of AES encryption.
\end{abstract}


\section{Introduction}
A biocryptography system with an encryption algorithm is an approach that combines biometric information with advanced encryption techniques to provide a high level of security for sensitive data \cite{bib1}. This system utilizes biometric data, such as fingerprints, facial recognition, or voice recognition, as a key to encrypt and decrypt data \cite{bib2}. The encryption algorithm provides an additional layer of security, making it extremely difficult for unauthorized individuals to access the protected information. Biocryptography with an encryption algorithm has a broad range of applications, from securing personal devices to protecting critical infrastructure systems. In this paper, we will explore the concepts of biocryptography and encryption algorithms, discuss the potential advantages and challenges of combining them, and provide a case study of a biocryptography system with an encryption algorithm in practice. Biometric characteristics including facial features, fingerprints, and iris patterns are unique to each individual and\newpage
\noindent and can serve as strong authentication factors in cryptographic systems \cite{bib3}. By using biometric data as cryptographic keys, biocryptography can provide secure access to information and prevent unauthorized access. One promising area of biocryptography is face-based authentication, which has several advantages over traditional authentication methods \cite{bib4}. Face recognition is a non-intrusive and user-friendly authentication method that can be easily integrated into existing systems \cite{bib5}. Additionally, facial recognition algorithms have become increasingly accurate and reliable, making it possible to use this biometric modality for high-security applications. Hence, biocryptography has the potential to revolutionize the field of security and privacy by providing a more secure and user-friendly means of authentication.
 This research proposes a biocryptosystem that uses facial recognition and AES encryption, with the key extracted from biological features, to provide a high level of security for sensitive data.
 By eliminating the need for passwords and relying solely on biometric authentication, this system
 reduces the possibility of illegal access to sensitive data. The proposed system ensures that it is user-friendly and simple to understand, especially for non
technical users. The biocryptosystem's performance standards are evaluated for precision, effectiveness, and security to guarantee its effectiveness and reliability. A user-friendly interface is created for the biocryptosystem to simplify its operation.

\section{Literature Survey}

Susanto and Ahmad\cite{bib6} address the computational complexity of pair-polar coordinate-based cancelable fingerprints. The process could yet be too slow for real-time applications and also does not discuss encryption applications. Zhang and Ding \cite{bib7} propose a method for encrypting digital images using the Advanced Encryption Standard (AES) algorithm for secure image transmission. Usman et al. \cite{bib8} implement the Secure Force (64-bit) encryption algorithm on a low-cost 8-bit microcontroller for a high level of security for data transmission while minimizing the cost of implementation. Li and Zhang \cite{bib9} use an efficient image encryption and decryption algorithm based on the combination of wavelet transform and chaos system. Ziang and Fu \cite{bib10} propose an image encryption scheme based on the Lorenz chaos system, which is a deterministic nonlinear system that exhibits chaotic behavior. The proposed scheme uses the Lorenz system to generate a chaotic key sequence, which is then used to encrypt the image data. P. N. et al.\cite{bib11} provides an application-specific integrated circuit (ASIC) implementation of the Rabbit stream cipher encryption algorithm for data security implementation. Wang and Ziang \cite{bib12} aim to provide a low-power and high-speed solution for data encryption in embedded systems using the 3-DES Encryption System with optimizing keys.

An improved design of the DES algorithm based on a symmetric encryption algorithm was done by Lihan and Yongzen \cite{bib13}, addressing the limitations of the DES algorithm, such as the small key space and vulnerability to brute-force attacks. Eskander et al. \cite{bib14} use a combination of a user's signature and face for authentication, with feature selection applied to both modalities to improve performance. Boosted feature selection is used to select a subset of features that are most discriminative for authentication. The scheme achieves a recognition rate of 95\% with a false acceptance rate of 1\%, indicating its potential for use in real-world biometric authentication applications. Gudkov and Ushmaev \cite{bib15} propose a user-dependent key extraction method from fingerprints using topological techniques. They leverage the topological structure of fingerprints to generate a unique cryptographic key for each user. By mapping the fingerprint to a topological space and extracting its homology groups, the method efficiently creates a secure key. Experimental results on a large fingerprint database validate the effectiveness of their approach.

Liu and Sarkar \cite{bib16} propose a method to improve gait recognition accuracy by normalizing gait dynamics. The approach extracts gait features using a spatio-temporal model, normalizes them with dynamic time warping, and trains a classifier on the aligned features. The authors demonstrate significant improvements in recognition accuracy on a large dataset with varied conditions, showcasing the method's effectiveness. Galbally and Marcel \cite{bib17} propose an image quality assessment (IQA) method for detecting fake biometrics in iris, fingerprint, and face recognition systems. The method analyzes structural and statistical properties of biometric images to distinguish between genuine and fake samples, enhancing security. Experimental evaluations on benchmark datasets demonstrate its effectiveness. Chandrasekaran \cite{bib18} presents a face recognition system using advanced neural networks to improve authentication security. Abdel-Ghaffar et al. \cite{bib19} propose a secure face recognition system using Gabor wavelet transform for feature extraction and SVM as a classifier. Tested on a dataset of 200 face images, the system achieved 94.5\% accuracy and showed robustness to occlusion and illumination changes. This system is promising for secure face recognition and could benefit from integrating additional biometric modalities. Zhao et al. \cite{bib20} explore a deep learning-based face image detection system for privacy and security in intelligent cloud platforms. The proposed framework uses deep learning algorithms for face detection and recognition, enhancing privacy by minimizing personal information exposure while maintaining effective security measures.

\section{Applied Methodology}
 There are many existing systems for face recognition that utilize various algorithms and techniques, including traditional computer vision approaches, deep learning models, and hybrid methods. Some popular face recognition systems include FaceNet, OpenFace, DeepFace, and VGGFace. Our method is based on the
 Histogram of Oriented Gradients (HOG) feature descriptor and Support Vector Machines (SVM) classifier. The HOG feature descriptor captures the gradient information of an image by dividing it into small cells and computing the orientation of gradients in each cell. The SVM classifier is then used to classify these features as belonging to a particular individual's face or not. The model is trained on a large dataset of faces to learn to recognize various facial features, such as the eyes, nose, and mouth, and to identify individuals based on these features. The model is also fine-tuned for specific applications, such as recognizing faces in low light or
 noisy environments, by adjusting the parameters of the HOG feature descriptor or the SVM classifier. Using
 these facial features, we are generating a 32-bit unique key for each user and then using this key in Advanced Encryption Standards (AES) algorithm to encrypt and decrypt the data.
The overall architecture diagram is shown in Fig. \ref{fig:face_auth_flow}
\begin{figure*}[t]
\centering
\begin{tikzpicture}[
    node distance=2cm,
    block/.style={rectangle, draw=teal!70!black, thick, fill=teal!10, text width=8em, text centered, minimum height=2.5em, rounded corners=3pt},
    smallblock/.style={rectangle, draw=blue!60!black, thick, fill=blue!10, text width=6em, text centered, minimum height=2.5em, rounded corners=3pt},
    network/.style={circle, draw=purple!60!black, thick, fill=purple!10, minimum size=2.5cm, align=center},
    arrow/.style={->, >=latex, thick, draw=gray!60!black}
]
    \node[smallblock] (username) {Enter\\Username};
    
    \node[smallblock, below left=0.7cm and 0.7cm of username] (newuser) {New User};
    \node[block, below=0.7cm of newuser] (enroll) {Face enrollment\\and storage\\in database};
    
    \node[smallblock, below right=0.7cm and 0.7cm of username] (existuser) {Existing User};
    \node[smallblock, below=0.7cm of existuser] (face) {User's Face};
    \node[block, below=0.7cm of face] (auth) {Face authen-\\tication using\\database};
    \node[block, below=0.7cm of auth] (key) {Key forma-\\tion/generation\\from face\\encoding};
    
    \node[network] at ($(key) + (0,-3cm)$) (network) {Open\\Network};
    
    \node[smallblock] at ($(network) + (-3.5cm,-0.5cm)$) (encrypt) {AES En-\\cryption};
    \node[smallblock] at ($(network) + (3.5cm,-0.5cm)$) (decrypt) {AES De-\\cryption};
    
    \node[left=1.2cm of encrypt, text=gray!40!black] (plain1) {Plain file};
    \node[above=0.07cm of network, text=gray!40!black] (cipher) {Cipher Image};
    \node[right=1.2cm of decrypt, text=gray!40!black] (plain2) {Plain file};
    
    \draw[arrow] (username) -- (newuser);
    \draw[arrow] (username) -- (existuser);
    \draw[arrow] (newuser) -- (enroll);
    \draw[arrow] (existuser) -- (face);
    \draw[arrow] (face) -- (auth);
    \draw[arrow] (auth) -- (key);
    
    \draw[arrow] (key.south west) -- (encrypt);
    \draw[arrow] (key.south east) -- (decrypt);
    
    \draw[arrow, dotted, draw=purple!60!black, thick] (encrypt) -- (network);
    \draw[arrow, dotted, draw=purple!60!black, thick] (network) -- (decrypt);
    
    \draw[arrow] (plain1) -- (encrypt);
    \draw[arrow] (decrypt) -- (plain2);
\end{tikzpicture}
\caption{Face Authentication-Based File Security System Flow Diagram.}
\label{fig:face_auth_flow}
\end{figure*}
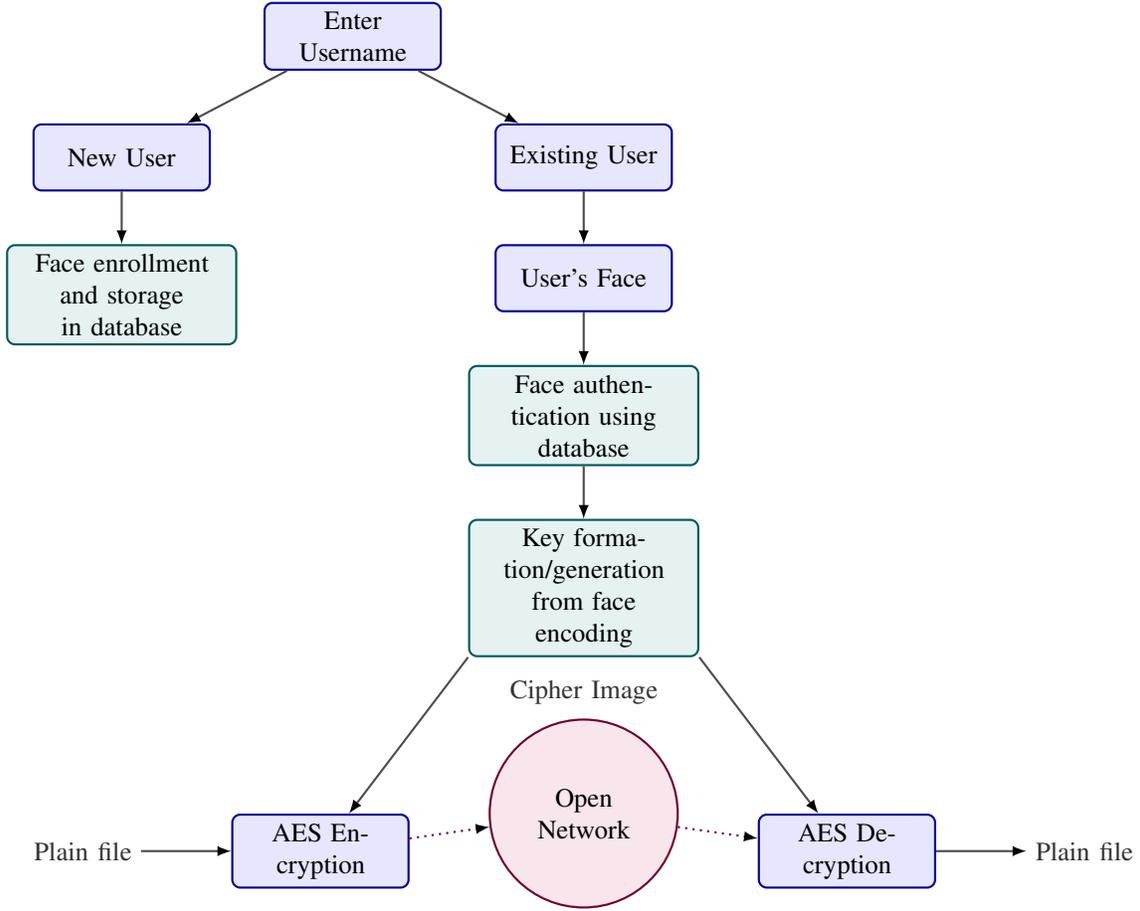

\subsection{Face enrollment and storage in database}
The image is captured using a webcam and after that, the following steps are performed:

\begin{itemize}
    \item \textbf{Image Preprocessing}: Convert the input image to grayscale. This step helps simplify the image and reduce computational complexity.
    \item \textbf{Sliding Window}: Iterate through the image using a sliding window approach. Move the window across the image in a grid-like fashion, typically with varying scales and aspect ratios.
    \item \textbf{Feature Extraction}: For each window position, extract the Histogram of Oriented Gradients (HOG) features. Divide the window into small cells and compute the gradient orientations within each cell. Concatenate these gradient orientations to form the HOG feature vector.
    \item \textbf{Feature Vector Representation}: The histograms of gradient orientations from all the cells within the window are concatenated to form a feature vector representation of the window. This feature vector captures the distinctive patterns of face-like structures.
    \item \textbf{SVM Classification}: Apply a trained Support Vector Machine (SVM) classifier to the extracted HOG features. The SVM classifier predicts whether the window contains a face or not based on the learned classification boundaries.
    \item \textbf{Thresholding}: Set a threshold for the SVM classifier's confidence score. If the confidence score exceeds the threshold, consider the window as a potential face region.
    \item \textbf{Non-maximum Suppression}: Perform non-maximum suppression to remove overlapping detections and keep only the most confident face regions. This step helps eliminate duplicate detections and improve the accuracy of the final results.
    \item \textbf{Bounding Box Drawing}: Draw bounding boxes around the final face regions. Each bounding box represents the location and size of a detected face in the input image.
\end{itemize}

\subsection{Face authentication using stored database}
The image is captured using a webcam and after that, the following steps are performed:

\begin{itemize}
    \item \textbf{Image Preprocessing}: Similar to the enrollment stage, the input image is preprocessed by converting it to grayscale to simplify the image and reduce computational complexity.
    \item \textbf{Sliding Window}: The sliding window approach is applied to iterate through the input image, just like in the enrollment stage. The window moves across the image in a grid-like fashion, at varying scales and aspect ratios.
    \item \textbf{Feature Extraction}: For each window position, the Histogram of Oriented Gradients (HOG) features are extracted. The window is divided into small cells, and the gradient orientations within each cell are computed. These gradient orientations are concatenated to form the HOG feature vector.
    \item \textbf{Feature Vector Representation}: The histograms of gradient orientations from all the cells within the window are concatenated to form a feature vector representing the window.
    \item \textbf{SVM Classification}: The trained Support Vector Machine (SVM) classifier from the enrollment stage is applied to the extracted HOG features. The SVM classifier predicts whether the window contains the face of the authorized user or not based on the learned classification boundaries.
    \item \textbf{Confidence Score and Thresholding}: The SVM classifier assigns a confidence score to each window. If the confidence score exceeds a predefined threshold, the window is considered a potential face region.
    \item \textbf{Non-maximum Suppression}: Non-maximum suppression is performed to remove overlapping detections and retain only the most confident face regions. This helps eliminate duplicate detections and enhance the accuracy of the final results.
    \item \textbf{Bounding Box Drawing}: Bounding boxes are drawn around the final face regions, indicating the location and size of the detected faces in the input image.
    \item \textbf{Comparison with Stored Face Templates}: The extracted face regions from the input image are compared with the stored face templates in the database. This comparison is done using similarity score between the feature vectors of the input face and the enrolled faces.
    \item \textbf{Authentication Decision}: Based on the comparison results, an authentication decision is made. If the similarity score or distance falls within an acceptable range or exceeds a predefined threshold, the face is considered a match, indicating successful authentication. Otherwise, the face is considered a mismatch, indicating that the person is not the authorized user.
\end{itemize}
\vspace{-1.2mm}
\subsection{Key formation/generation using face encoding}
\begin{itemize}
    \item \textbf{Face Encoding}: Face encoding is performed using the HOG with SVM approach using the existing user’s registered image. This process extracts facial features and represents them as a numerical vector, often of higher dimensions.
    \item \textbf{Vector Transformation}: From the face encoding vector, each value is converted to a string.
    \item \textbf{Selecting Key Values}: From the transformed vector to string, the first 32 values are selected as we are using a 32-byte AES architecture for further encryption and decryption.
    \item \textbf{Key Representation}: The selected key is typically represented in a suitable format for storage (i.e., byte string representation).
\end{itemize}

\subsection{Bio-key based AES Encryption}
\begin{itemize}
    \item The \texttt{encrypt} function accepts plaintext (\(P\)) and a 32-byte cryptographic key (\(K\)) as input.
    
    \item If the length of \(P\) (\(L_P\)) is not a multiple of the AES block size (16 bytes), null-byte padding (\(0x00\)) is applied to ensure:
    \[
    L_P \mod 16 = 0
    \]

    \item A 16-byte random Initialization Vector (\(IV\)) is generated to introduce randomness and prevent identical ciphertexts for identical plaintexts.
    
    \item An AES cipher object is initialized in Cipher Block Chaining (CBC) mode using \(K\) and \(IV\).
    
    \item The padded plaintext is encrypted block-by-block in CBC mode:
    \[
    C_i = AES_K(P_i \oplus C_{i-1}), \quad \text{where } C_0 = IV
    \]
    
    \item The final encrypted output (\(C\)) is constructed by prepending the \(IV\) to the concatenated ciphertext blocks:
    \[
    C = IV \parallel C_1 \parallel C_2 \parallel \ldots \parallel C_n
    \]
\end{itemize}

\subsection{Bio-key based AES Decryption}
\begin{itemize}
    \item The \texttt{decrypt} function accepts the ciphertext (\(C\)) and the 32-byte cryptographic key (\(K\)) as input.
    
    \item The IV is extracted from the beginning of the ciphertext:
    \[
    IV = C_0
    \]
    
    \item An AES cipher object is initialized in Cipher Block Chaining (CBC) mode using \(K\) and the extracted \(IV\).
    
    \item The ciphertext (excluding the IV) is decrypted block-by-block in CBC mode:
    \[
    P_i = AES_K^{-1}(C_i) \oplus C_{i-1}, \quad \text{where} \quad C_0 = IV
    \]
    where \(AES_K^{-1}\) represents the decryption operation with the key \(K\).
    
    \item The decrypted output is obtained, which includes the padded plaintext.
    
    \item Any trailing null bytes are stripped from the decrypted output to retrieve the original plaintext:
    \[
    P = \text{strip}(P_{\text{padded}})
    \]
\end{itemize}

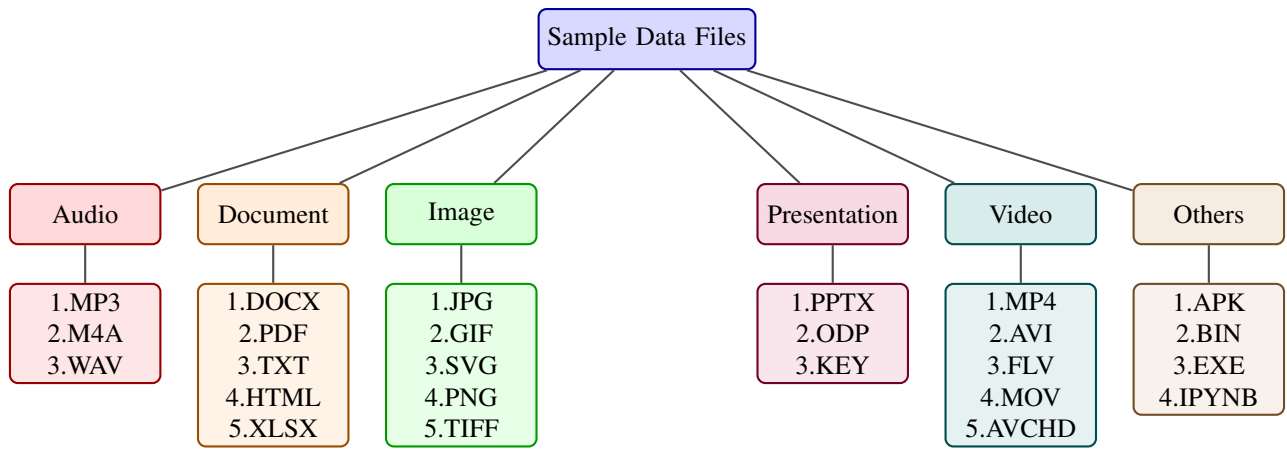
\begin{figure*}[t]
\centering
\begin{tikzpicture}[
    level distance=2cm,
    sibling distance=3cm,
    every node/.style={
        draw=gray!60!black,
        thick,
        rectangle,
        minimum width=2cm,
        minimum height=0.8cm,
        align=center,
        rounded corners=3pt
    }
]
\node[fill=blue!15, draw=blue!60!black] (root) {Sample Data Files};

\node[fill=red!15, draw=red!60!black] (audio) [below left=1.5cm and 5cm of root] {Audio};
\node[fill=orange!15, draw=orange!60!black] (document) [below left=1.5cm and 2.5cm of root] {Document};
\node[fill=green!15, draw=green!60!black] (image) [below left=1.5cm and 0cm of root] {Image};
\node[fill=purple!15, draw=purple!60!black] (presentation) [below right=1.5cm and 0cm of root] {Presentation};
\node[fill=teal!15, draw=teal!60!black] (video) [below right=1.5cm and 2.5cm of root] {Video};
\node[fill=brown!15, draw=brown!60!black] (others) [below right=1.5cm and 5cm of root] {Others};

\foreach \i in {audio,document,image,presentation,video,others}
    \draw[gray!60!black, thick] (root) -- (\i);

\node[fill=red!10, draw=red!60!black] (audio-formats) [below=0.5cm of audio] {1.MP3\\2.M4A\\3.WAV};
\draw[gray!60!black, thick] (audio) -- (audio-formats);

\node[fill=orange!10, draw=orange!60!black] (document-formats) [below=0.5cm of document] {1.DOCX\\2.PDF\\3.TXT\\4.HTML\\5.XLSX};
\draw[gray!60!black, thick] (document) -- (document-formats);

\node[fill=green!10, draw=green!60!black] (image-formats) [below=0.5cm of image] {1.JPG\\2.GIF\\3.SVG\\4.PNG\\5.TIFF};
\draw[gray!60!black, thick] (image) -- (image-formats);

\node[fill=purple!10, draw=purple!60!black] (presentation-formats) [below=0.5cm of presentation] {1.PPTX\\2.ODP\\3.KEY};
\draw[gray!60!black, thick] (presentation) -- (presentation-formats);

\node[fill=teal!10, draw=teal!60!black] (video-formats) [below=0.5cm of video] {1.MP4\\2.AVI\\3.FLV\\4.MOV\\5.AVCHD};
\draw[gray!60!black, thick] (video) -- (video-formats);

\node[fill=brown!10, draw=brown!60!black] (other-formats) [below=0.5cm of others] {1.APK\\2.BIN\\3.EXE\\4.IPYNB};
\draw[gray!60!black, thick] (others) -- (other-formats);

\end{tikzpicture}
\caption{Classification of Sample Data Files into Media Categories.}
\label{fig:face_auth_flow}
\end{figure*}

\section{Results and Discussion}

\subsection{Sample Data}
For experiment and performance analysis of the designed system, we have used a total of 25 files of different extensions and categorized them into media types as shown in Fig. \ref{fig:face_auth_flow}.

\subsection{Correlation Analysis}
The correlation coefficient is a statistical measure that quantifies the linear relationship between two datasets \cite{bibx}. It can provide insights into the similarity or dissimilarity between the plain and encrypted files.

In the context of evaluating encryption strength, the correlation coefficient is relevant as it indicates the degree of linear relationship between the original and encrypted data. A higher positive correlation coefficient suggests a stronger linear relationship or similarity between the two files. This can be a potential vulnerability as it may reveal patterns or information about the original data in the encrypted file.

Conversely, a lower correlation coefficient or a negative correlation coefficient indicates a weaker linear relationship or even an inverse relationship between the two datasets. In the case of encryption, a low or negative correlation coefficient is desirable as it suggests that the encryption process has introduced a significant level of randomness and made it difficult to discern any meaningful information about the original data from the encrypted data. A lower or negative correlation coefficient indicates a higher level of encryption strength, as it implies that the encrypted data exhibits a high degree of randomness and lacks patterns that could reveal information about the original data.

Therefore, when assessing encryption strength, a low or negative correlation coefficient is preferred, indicating a weak or no linear relationship between the plain and encrypted files and suggesting a higher level of data protection.
The correlation coefficient is given by:

\[
r_{xy} = \frac{\sum_{i=1}^{n}(x_i - \bar{x})(y_i - \bar{y})}{\sqrt{\sum_{i=1}^{n}(x_i - \bar{x})^2} \sqrt{\sum_{i=1}^{n}(y_i - \bar{y})^2}}
\]

  where \(x_i\) is the \(i\)-th bit of the plain file, \(\bar{x}\) is the mean of all bits of the plain file, \(y_i\) is the \(i\)-th bit of the encrypted file, \(\bar{y}\) is the mean of all bits of the encrypted file, and \(n\) is the total number of samples.

The range of values for the correlation coefficient \( r_{xy} \) is from \(-1\) to \(1\), i.e.,

In the correlation analysis performed by us in the Table \ref{tab:efficiency_measures}, we can see that almost all types of files have correlation coefficient are less than 0.05 or are negative. Hence, this data encryption algorithm is highly strong. 

\subsection{Shannon Entropy Analysis}
Shannon entropy, named after Claude Shannon, is a concept used in information theory to measure the average amount of information or uncertainty in a set of data \cite{bib21}. It can be used as a metric to assess the strength of encryption. Shannon entropy value provides an indication of the randomness and unpredictability of the encrypted data. A higher entropy value indicates higher randomness and increased encryption strength. Conversely, a lower entropy value suggests patterns or predictability in the encrypted data, indicating potential weaknesses in the encryption. 

The encryption strength is measured using Shannon entropy by first selecting a representative sample of the encrypted data. Next, the frequency distribution of each symbol or character in the sample is calculated. The probability of each symbol is then computed by dividing its frequency by the total number of symbols. Finally, Shannon entropy is calculated using the standard formula given below:
 \[
H(X) = -\sum_{i=1}^n p(x_i) \log_b p(x_i)
\]
where, \(x_i\) represents the \(i\)-th symbol, \(p(x_i)\) is the probability of the occurrence of an individual symbol \(x_i\) in the data, and \(n\) is the total number of unique symbols in the data.

\begin{table*}[htbp]
\centering
\renewcommand{\arraystretch}{1.3}
\setlength{\tabcolsep}{5pt}
\begin{tabular*}{\textwidth}{@{\extracolsep{\fill}}|l|c|c|c|c|c|}
\hline
\textbf{Files} & \textbf{Correlation} & \textbf{Shannon Entropy Plain} & \textbf{Shannon Entropy Cipher} & \textbf{Normalized Hamming Distance} & \textbf{Avalanche Effect (in \%)} \\
\hline
MP3 & -0.000039032 & 7.476443127 & 7.619993924 & 0.49993309 & 49.9994066 \\
M4A & -0.000417714 & 4.369103925 & 7.636474301 & 0.500014403 & 50.0071549 \\
WAV & -0.000813543 & 3.887155306 & 7.634233576 & 0.499972689 & 49.9904273 \\
DOCX & -0.000327663 & 7.646910535 & 7.639741906 & 0.500253356 & 49.9852590 \\
HTML & 0.049166343 & 4.685769328 & 7.061617375 & 0.496635095 & 48.6816406 \\
PDF & -0.000155853 & 7.464564052 & 7.618633544 & 0.499886012 & 50.0285371 \\
TEXT & 0.019436802 & 5.505292686 & 7.437564883 & 0.500319157 & 49.693145 \\
XLSX & -0.001176861 & 5.859034131 & 7.421994201 & 0.500522037 & 50.2050089 \\
GIF & -0.000899482 & 7.617688861 & 7.596312484 & 0.499734286 & 49.9872537 \\
JPG & -0.001086416 & 7.469949329 & 7.595499825 & 0.500516713 & 49.9942359 \\
PNG & 0.003620823 & 7.590157683 & 7.611722336 & 0.499951895 & 50.0067973 \\
SVG & -0.00083704 & 4.031431828 & 7.541224011 & 0.49932164 & 49.9392903 \\
TIFF & -0.000168505 & 8.440207014 & 7.61661058 & 0.500161339 & 50.0091104 \\
KEY & -0.000915817 & 6.956186376 & 7.593541712 & 0.500297172 & 50.0042947 \\
PPTX & -0.000398026 & 7.452324086 & 7.631706223 & 0.499942322 & 50.000741 \\
ODP & -0.000559883 & 7.509882306 & 7.617472309 & 0.50016724 & 49.9901331 \\
AVCHD & -0.000520346 & 7.565098635 & 7.637302042 & 0.500064329 & 49.9961774 \\
AVI & -0.00013142 & 5.28734789 & 7.616158108 & 0.5000433 & 49.9792113 \\
FLV & -0.000257436 & 7.624811313 & 7.636109029 & 0.499985164 & 50.0088053 \\
MOV & -0.000568506 & 5.233715571 & 7.625062336 & 0.500251231 & 50.0136696 \\
MP4 & -0.000165741 & 7.426341212 & 7.638600787 & 0.500018749 & 49.9983425 \\
APK & 0.000125946 & 7.650620934 & 7.633420708 & 0.499975041 & 50.0036145 \\
BIN & -0.000348078 & 5.326272874 & 7.514698302 & 0.498799642 & 50.0107458 \\
EXE & -0.000263785 & 7.001071393 & 7.642106235 & 0.500071483 & 50.0082617 \\
IPYNB & -0.016073869 & 4.798464084 & 7.374466006 & 0.50057233 & 49.8566938 \\
\hline
\end{tabular*}
\caption{Efficiency Measures of each file.}
\label{tab:efficiency_measures}
\end{table*}

The logarithmic base (b=2) used in the Shannon entropy formula is derived from the fundamental nature of binary information representation. In binary representation, information is typically represented using bits (binary digits) with two possible values: 0 or 1. The logarithmic base (b=2) aligns with this binary nature, as it allows for the measurement of information in terms of bits.
In the table below we have calculated the Shannon entropy of the plain file as well as the encrypted file.

The Shannon entropy of an encrypted file can have a maximum value of 8, as the maximum entropy per byte symbol is calculated as the logarithmic base (b=2) of the number of possible bits symbols (we have used AES 256 bits), and log2(256) is 8.

The Shannon entropy file can have a maximum value of more than 8 depending on the high randomness, complexity, and level of detail in the file. Images with more varied pixel values, textures, and colors tend to have higher entropy values, as demonstrated by the efficiency table, where TIFF has an entropy value of 8.440207014.

\subsection{Normalized Hamming Distance}
The normalized Hamming distance is a measure that can be used to assess the efficiency or effectiveness of an encryption algorithm \cite{bib22}. It quantifies the similarity or dissimilarity between two binary sequences, such as here the original data and the encrypted data. A higher normalized distance indicates a higher level of encryption efficiency, suggesting that the encryption algorithm has effectively scrambled the data and introduced a high degree of randomness. On the other hand, a lower normalized distance could indicate potential weaknesses in the encryption algorithm, as it suggests a greater similarity or predictability between the original and encrypted data.

To achieve a balance between security and computational complexity, it is important to consider a normalized Hamming distance that ensures encrypted data is moderately dissimilar to the original data. This provides a reasonable level of security while maintaining efficiency, depending on the system's specific requirements and data sensitivity.

In our case, we achieved a normalized Hamming distance of approximately 0.5 for various file types. This strikes a balance by ensuring the encrypted data is sufficiently distinct from the original data without incurring excessive computational overhead. A normalized distance closer to 1 would provide higher security but at greater computational cost, while a distance near 0 would indicate insufficient encryption.

\[
d_{\text{norm}} = \frac{d}{n}
\]
where d is the Hamming distance and n is the total number of bits. \\
Range of values of \(d_{\text{norm}}\): [0,1].

\subsection{Avalanche Effect Analysis}
The avalanche effect is a desirable property in encryption algorithms that ensures that even a small change in the input or key results in significant changes in the output or ciphertext \cite{bib23}. It is one of the criteria used to evaluate encryption strength and the resilience of an encryption algorithm against attempts to deduce information about the original data.

To evaluate our encryption algorithm using the avalanche effect, we have compared the ciphertexts of the original file and the ciphertext of the modified version of the original file. We have modified 50\% of bytes of the original file at its random positions by using bitwise XOR (\(\oplus\)) between the selected byte and the left bit-shifting operation \(1 \ll (\text{position} \bmod 8)\). After that, the avalanche effect percentage (\%) is calculated as follows:

\[
\boxed{
\frac{\text{Number of flipped bits in ciphertext}}{\text{Total number of bits in ciphertext}} \times 100
}
\]

\[
\text{Avalanche Effect (in \%)} = \left(\frac{d}{n}\right) \times 100
\]
where
d is the number of differentiating bits between two files
n is total number of bits.
Range of values: [0\% to 100\%]

\section{Conclusion}
The proposed system based on the Histogram of Oriented Gradients (HOG) feature descriptor and Support Vector Machines (SVM) classifier shows promising results in facial recognition and encryption of data. However, there is still scope for further improvement and development. One area for future research could be to explore the use of deep learning techniques, such as convolutional neural networks (CNNs), for facial recognition. Another area of future research could be to improve the performance of the system in low-light or noisy environments by incorporating additional features, such as color or texture in the feature descriptor or by using denoising techniques to preprocess the images. Furthermore, the proposed system can be extended to other applications, such as surveillance, access control, and identity verification. The biocryptosystem can also be utilized for secure file sharing between individuals or within a company. The biocryptosystem is also capable of protecting
online transactions like e-commerce and banking. Face recognition technology can be used to verify users' identities, and the AES encryption technique can be used to encode and decrypt user transactions to guard against fraud and illegal access. The biocryptosystem can be used to control access to protected digital or physical locations. The biocryptosystem can also be used to protect sensitive patient data by securing medical records. Face recognition technology can be used to verify patients, and the AES encryption method can be used to encrypt and decrypt patient records to prevent illegal access. 

Overall, the proposed system has great potential in the field of facial recognition and data encryption and has many potential applications and provides a reliable and effective way to safeguard sensitive data. This innovative integration of facial biometrics and AES encryption lays the groundwork for future advancements in secure digital ecosystems.


\newpage
\begin{thebibliography}{00}


\bibitem{bib1}
K. Xi and J. Hu, "Bio-Cryptography," in \textit{Handbook of Information and Communication Security}, P. Stavroulakis and M. Stamp, Eds. Berlin, Heidelberg: Springer Berlin Heidelberg, 2010, pp. 129–157.

\bibitem{bib2}
Z. Khalid, M. Rizwan, A. Shabbir, M. Shabbir, F. Ahmad, and J. Manzoor, ``Cloud server security using bio-cryptography,'' in \textit{International Journal of Advanced Computer Science and Applications}, vol. 10, no. 3, 2019, pp. 166--172.

\bibitem{bib3}
R. Alrawili, A. A. S. AlQahtani, and M. K. Khan, ``Comprehensive survey: Biometric user authentication application, evaluation, and discussion,'' \textit{Computers and Electrical Engineering}, vol. 119, Part A, 2024, Art. no. 109485, ISSN 0045-7906.

\bibitem{bib4}
S. Aanjanadevi, V. Palanisamy, S. Aanjankumar, and S. Poonkuntran, "A Secure Authenticated Bio-cryptosystem Using Face Attribute Based on Fuzzy Extractor," in \textit{New Trends in Computational Vision and Bio-inspired Computing}, S. Smys, A. M. Iliyasu, R. Bestak, and F. Shi, Eds. Cham: Springer, 2020.


\bibitem{bib5}
A. K. Jain and A. Kumar, ``Biometric recognition: An overview,'' in \textit{Second Generation Biometrics: The Ethical, Legal and Social Context}, E. Mordini and D. Tzovaras, Eds. Dordrecht: Springer, 2012, vol. 11, pp. 35–56.

\bibitem{bib6}
L. Susanto and T. Ahmad, ``Reducing Computational Cost of Pair-Polar Coordinate-based Cancelable Fingerprint Template Matching,'' in \textit{2020 3rd International Conference on Computer and Informatics Engineering (IC2IE)}, Yogyakarta, Indonesia, 2020, pp. 31--36.

\bibitem{bib7}
Q. Zhang and Q. Ding, ``Digital Image Encryption Based on Advanced Encryption Standard (AES),'' in \textit{2015 Fifth International Conference on Instrumentation and Measurement, Computer, Communication and Control (IMCCC)}, Qinhuangdao, China, 2015, pp. 1218--1221.

\bibitem{bib8}
M. Usman, S. Z. -U. -A. Abidi, M. H. S. Siddiqui, and M. S. Ibrahim, ``Implementation of Secure Force (64-bit) on low cost 8-bit microcontroller,'' in \textit{2016 International Conference on Open Source Systems \& Technologies (ICOSST)}, Lahore, Pakistan, 2016, pp. 102--105.

\bibitem{bib9}
X. Li and Y. Zhang, ``Digital image encryption and decryption algorithm based on wavelet transform and chaos system,'' in \textit{2016 IEEE Advanced Information Management, Communicates, Electronic and Automation Control Conference (IMCEC)}, Xi'an, China, 2016, pp. 253--257.

\bibitem{bib10}
H. -y. Jiang and C. Fu, ``An Image Encryption Scheme Based on Lorenz Chaos System,'' in \textit{2008 Fourth International Conference on Natural Computation}, Jinan, China, 2008, pp. 600--604.

\bibitem{bib11}
P. N., C. S., and S. M. Rehman, ``ASIC Implementation of Rabbit Stream Cipher Encryption for Data,'' in \textit{2019 IEEE International WIE Conference on Electrical and Computer Engineering (WIECON-ECE)}, Bangalore, India, 2019, pp. 1--4.

\bibitem{bib12}
L. Wang and G. Jiang, ``The Design of 3-DES Encryption System Using Optimizing Keys,'' in \textit{2019 China-Qatar International Workshop on Artificial Intelligence and Applications to Intelligent Manufacturing (AIAIM)}, Doha, Qatar, 2019, pp. 56--58.

\bibitem{bib13}
W. Yihan and L. Yongzhen, ``Improved Design of DES Algorithm Based on Symmetric Encryption Algorithm,'' in \textit{2021 IEEE International Conference on Power Electronics, Computer Applications (ICPECA)}, Shenyang, China, 2021, pp. 220--223.

\bibitem{bib14}
G. S. Eskander, R. Sabourin, and E. Granger, ``Signature based Fuzzy Vaults with Boosted Feature Selection,'' in \textit{2011 IEEE Workshop on Computational Intelligence in Biometrics and Identity Management (CIBIM)}, Paris, France, 2011, pp. 131--138.

\bibitem{bib15}
V. Y. Gudkov and O. Ushmaev, ``A Topologic Approach to User-Dependent Key Extraction from Fingerprints,'' in \textit{2010 20th International Conference on Pattern Recognition}, Istanbul, Turkey, 2010, pp. 1281--1284.

\bibitem{bib16}
Z. Liu and S. Sarkar, ``Improved gait recognition by gait dynamics normalization,'' \textit{IEEE Transactions on Pattern Analysis and Machine Intelligence}, vol. 28, no. 6, pp. 863--876, Jun. 2006.


\bibitem{bib17}
J. Galbally, S. Marcel, and J. Fierrez, ``Image Quality Assessment for Fake Biometric Detection: Application to Iris, Fingerprint, and Face Recognition,'' \textit{IEEE Transactions on Image Processing}, vol. 23, no. 2, pp. 710--724, Feb. 2014.

\bibitem{bib18}
R. Chandrasekaran, ``Security Assurance Using Face Recognition \& Detection System Based On Neural Networks,'' in \textit{2005 International Conference on Neural Networks and Brain}, Beijing, China, 2005, pp. 1119--1106.

\bibitem{bib19}
E. A. Abdel-Ghaffar, M. E. Allam, H. A. K. Mansour, and M. A. Abo-Alsoud, ``A secure face recognition system,'' in \textit{2008 International Conference on Computer Engineering \& Systems}, Cairo, Egypt, 2008, pp. 95--100.

\bibitem{bib20}
X. Zhao, S. Lin, X. Chen et al., ``Application of face image detection based on deep learning in privacy security of intelligent cloud platform,'' \textit{Multimedia Tools and Applications}, vol. 79, pp. 16707--16718, 2020.
\bibitem{bibx}
P. Schober, C. Boer, and L.A. Schwarte, ``Correlation Coefficients: Appropriate Use and Interpretation,'' in \textit{Anesthesia \& Analgesia}, vol. 126, no. 5, pp. 1763--1768, May 2018. \href{10.1213/ANE.0000000000002864}.

\bibitem{bib21}
T.O. Kvålseth, ``On the Measurement of Randomness (Uncertainty): A More Informative Entropy,'' in \textit{Entropy}, vol. 18, no. 5, 2016, p. 159. \href{10.3390/e18050159}.
\bibitem{bib22}
R. Kulkarni and A. Namboodiri, ``Secure hamming distance based biometric authentication,'' in \textit{2013 International Conference on Biometrics (ICB)}, Madrid, Spain, 2013, pp. 1--6.
\bibitem{bib23}
O. C. Abikoye, A. D. Haruna, A. Abubakar, N. O. Akande, and E. O. Asani, ``Modified Advanced Encryption Standard Algorithm for Information Security,'' \textit{Symmetry}, vol. 11, Art. no. 1484, 2019.



\end{thebibliography}
\end{document}